\documentclass[twocolumn,aps,prb,floatfix,showpacs,amsmath,amssymb]{revtex4-1}

\usepackage{epsfig}
\usepackage{graphicx}
\usepackage{dcolumn}
\usepackage{bm}
\usepackage[colorlinks=true,dvipdfm]{hyperref}
\begin{document}
\title{Universal short-time quantum critical dynamics in imaginary time}
\author{Shuai Yin} \email{zsuyinshuai@163.com}
\author{Peizhi Mai}
\author{Fan Zhong}  \email{stszf@mail.sysu.edu.cn}

\affiliation{State Key Laboratory of Optoelectronic Materials and
Technologies, School of Physics and Engineering, Sun Yat-sen
University, Guangzhou 510275, People's Republic of China}

\begin{abstract}
We propose a scaling theory for the universal imaginary-time quantum critical dynamics for both short times and long times. We discover that there exists a universal critical initial slip related to a small initial order parameter $M_0$. In this stage, the order parameter $M$ increases with the imaginary time $\tau$ as $M\propto M_0\tau^\theta$ with a universal initial slip exponent $\theta$. For the one-dimensional transverse-field Ising model, we estimate $\theta$ to be $0.373$, which is markedly distinct from its classical counterpart. Apart from the local order parameter, we also show that the entanglement entropy exhibits universal behavior in the short-time region. As the critical exponents in the early stage and in equilibrium are identical, we apply the short-time dynamics method to determine quantum critical properties. The method is generally applicable in both the Landau-Ginzburg-Wilson paradigm and topological phase transitions.
\end{abstract}


\maketitle

\date{\today}
\section{\label{introduction}Introduction}
Universal properties exhibited in continuous quantum phase transitions are controlled by low energy levels. \cite{sachdev,sondhi} These universal properties are often described by critical exponents which are not sensitive to the microscopic information of a system. It is well known that the universal static properties of a $d$ dimensional quantum system correspond to those of a $d+1$ dimensional classical system. \cite{sachdev,sondhi} This correspondence can be seen from the imaginary-time path integral. \cite{sachdev,sondhi} However, there is no direct mapping between classical and quantum dynamics. It is thus expected that dynamic quantum criticality has some unique properties. \cite{Dziarmaga,polrmp} In fact, in contrast to the classical one, quantum critical dynamics cannot be separated from the statics. \cite{sachdev,sondhi} Therefore, dynamics is pivotal to understand quantum phase transitions. Besides its fundamental interest, it may lead to better control in adiabatic quantum computations. \cite{Dziarmaga}

A lot of efforts have indeed been devoted to understand dynamic quantum criticality. Experimental advances have provided effective platforms to manipulate and observe accurately dynamic quantum critical behavior. \cite{Greiner,mei,ulm,pyka} When a system is driven across its quantum critical point, the Kibble-Zurek mechanism~\cite{ulm,pyka,kibble1,zurek1,zurekq,Dziarmaga,polrmp} predicts the scaling of the density of defects generated due to the breakdown of adiabaticity in the impulse region. Finite-time scaling just fits the nonequilibrium dynamics in this region in which the driving time is shorter than the reaction time and accounts for the scaling found in the region. \cite{dengd,dengde,dengko,Yin} When a system is subjected to a sudden quench near its quantum critical point, universal properties have also been found at long times when properties of the ground state dominate. \cite{degrandi,deng,campos,iyer,karrasch,Rossini,Foini,Calabrese}

Here, we shall focus on universal behavior in short times after a sudden quench. Apparently, this short-time dynamics depends on the realization of the initial states. When a system is suddenly quenched off its critical point, universal short-time behavior has been shown to exist, \cite{polkovprl} because at short times, dynamic behavior is still controlled by the energy levels near the ground state. A universal scaling relation connecting the short-time dynamics to the long-time dynamics was also found. \cite{polkovprl} In contrast, when a parameter is initially set at a noncritical value and then suddenly quenched to the critical point, does universal short-time behavior exist either?

This question is partly motivated by the discovery and application of universal short-time dynamics in the relaxational critical process in classical phase transitions.\cite{Janssen,Li2,Zheng} When a classical system is suddenly quenched from very high temperature with a small magnetization $M_0$ and vanishing correlation to the critical region, it has been found that there exists a new short-time stage showing universal behavior.\cite{Janssen} Right after the quench, the evolution of the system is dominated by microscopic details and so no universal scaling behavior exists. At long times, the system comes to the period characterized by the familiar power-law decay. In this region, the order parameter varies with time as $M\sim t^{-\beta/\nu z}$, where $\beta$ and $\nu$ are static critical exponents defined by $M\sim (T-T_c)^\beta$ ($T$ is the temperature and $T_c$ the critical temperature) and the correlation length $\xi\sim (T-T_c)^{-\nu}$, respectively, in the equilibrium situation and $z$ is the dynamic critical exponent defined by $\zeta\sim \xi^z$ with $\zeta$ being the correlation time. In between, the system enters the short-time stage characterized by a ``critical initial slip",\cite{Janssen} where $M$ increases surprisingly as $M\sim M_0 t^\theta$ with a new universal initial-slip exponent $\theta$. As the short-time dynamics overcomes the difficulties induced by critical slowing down, it has become a powerful method to determine the critical properties.\cite{Li2,Zheng}

Some efforts have been attempted to extend this kind of short-time dynamics into quantum situations. \cite{zhengq,Santos} The results obtained cannot, however, fully reflect the quantum properties. For instance, in the one dimensional (1D) transverse field Ising model, the dynamic exponent obtained based on the Metropolis dynamics of quantum Monte Carlo is $z=1.883(7)$,\cite{Santos} which apparently disagrees with the exact result $z=1$.\cite{sachdev,sondhi} This is because the Metropolis dynamics cannot catch the properties of quantum dynamics.\cite{zhengq,binder} Therefore, in contrast to the Kibble-Zurek mechanism which has been proved to be applicable in both quantum and classical dynamics,\cite{Dziarmaga,polrmp} it is still unclear whether similar short-time dynamics is applicable to quantum situations or not.

Study of real-time quantum dynamics is hindered by lack of effective numerical methods. For example, powerful Monte Carlo methods fail to grasp the unitary properties of quantum dynamics because of the sign problem.\cite{binder} It is also a challenge to calculate real-time evolution in density-matrix renormalization-group methods.\cite{scholl1} Fortunately, recent studies of imaginary-time quantum dynamics shed some light on this problem.\cite{degrandi2} Firstly, imaginary-time evolutions are readily realized in quantum Monte Carlo~\cite{degrandi2,cwliu,degrandi3} and time-dependent density-matrix renormalization-group methods \cite{vidal1,vidal,jiang} and are in fact a popular method in determining the ground state. \cite{vidal1,vidal,jiang,justin} Secondly, some properties are shared in both real-time and imaginary-time evolutions. For example, a critical quench in the imaginary-time direction confirms the Kibble-Zurek mechanism predicted in real-time dynamics.\cite{degrandi2} Another example is that the universal real-time dynamic behavior of the kinetic energy per length in the Tomonaga-Luttinger model is the same as its imaginary-time evolution except the early oscillations in the real-time situation.\cite{karrasch} So, imaginary-time evolutions have demonstrated their power in the study of quantum dynamics, although analytic continuation does not always work.\cite{altland}

In this paper, we explore primarily the universal properties of short-time quantum critical dynamics in the imaginary time direction. A scaling theory is proposed which can account for both the short-time and long-time universal properties. The initial state is chosen to be a direct product state with a vanishing correlation length. Such a state has no entanglement and can have an arbitrarily small magnetization $M_0$. Similar to the classical situation, a universal critical initial slip characterized by $M\propto M_0\tau^\theta$ for small $M_0$ appears after the non-universal transient in imaginary time $\tau$. However, $\theta=0.373$ for the 1D transverse field Ising model, in sharp difference from its classical counterpart, which is $0.191(1)$~\cite{Li3,Zheng1,zhengtri,Zheng,Albano} for the 2D classical Ising model. Thus $\theta$ is a new quantum dynamic exponent. Besides $\theta$, the universal short-time dynamics is also characterized by critical exponents that are identical with those describing the long-time behavior.

An effective method based on the short imaginary-time quantum-critical dynamics (SITQCD) is then developed to determine the quantum critical properties. An apparent advantage is that this method circumvents critical slowing down which also manifests in quantum phase transitions. We shall show that this method is applicable to phase transitions both in the Landau-Ginzburg-Wilson paradigm and of topological nature.

The rest of the paper is organized as follows. After introducing briefly the imaginary-time evolution in Sec.~\ref{ITE}, we propose a scaling theory to describe the universal imaginary-time quantum critical dynamics both for short times and long times in Sec.~\ref{scaling}. In Sec.~\ref{theta}, we verify the scaling theory using mainly the 1D transverse field Ising model as an example. A mean-field theory is developed which conforms with the scaling theory but yields no initial slip. Numerical solutions of the models are then utilized to verify the universal short-time behavior. The initial-slip exponent $\theta$ is determined and the full scaling forms are verified for both short-time and long-time behaviors. These scaling forms are then employed back as an effective method to determine the critical properties in Sec.~\ref{application}. We first benchmark the method with the $1$D Ising model. Then we apply it to determine the critical properties of the topological phase transition in the anisotropic spin-$1$ Heisenberg model. Short real-time dynamics is discussed in Sec. \ref{real} and a summary is given in Sec.~\ref{summary}.

\section{\label{ITE}Imaginary time evolution and long time behavior}
In this section, we briefly review the imaginary time evolution of a quantum system described by a Hamiltonian $H$. Evolution of a quantum state $|\psi(t)\rangle$ is given by the time-dependent Schr\"{o}dinger equation with an initial wave function $|\psi_0 \rangle=|\psi(0) \rangle$. To describe the imaginary-time evolution, $t$ is replaced by $-i\tau$ and the Plank constant is set to 1, and Schr\"{o}dinger's equation becomes \cite{justin}
\begin{equation}
{\frac{\partial}{\partial\tau} |\psi(\tau) \rangle}=-H|\psi(\tau)\rangle, \label{scheq}
\end{equation}
with the normalization condition $\langle \psi(\tau)|\psi(\tau) \rangle=1$. Its formal solution is
\begin{equation}
|\psi(\tau) \rangle=Z\textrm{exp}(-H\tau)|\psi_0 \rangle, \label{sscheq}
\end{equation}
where $Z=1/\|\textrm{exp}(-H\tau)|\psi_0\rangle\|$ is the normalization factor and $\|\centerdot\|$ denotes a modulo operation.

The long time behavior of the evolution can be recognized from Eq.~(\ref{sscheq}). For an initial state with nonzero projection on the ground state, $|\psi(\tau) \rangle$ can be explicitly calculated by expanding the initial state $|\psi_0 \rangle$ in the energy representation
\begin{eqnarray}
\begin{split}
|\psi(\tau) \rangle&= Z\sum_i c_i \textrm{e}^{-E_i\tau}|E_i\rangle \\&= Z\textrm{e}^{-E_0\tau}\sum_i c_i \textrm{e}^{-(E_i-E_0)\tau}|E_i\rangle \\&\sim c_0 |E_0\rangle+c_1 \textrm{e}^{-\Delta \tau}|E_1\rangle, \label{sscheq1}
\end{split}
\end{eqnarray}
where $c_i=\langle E_i|\psi_0\rangle$, $|E_i\rangle$ is the $i$th eigenstate of $H$ with the eigenvalue $E_i$ ordered with $i$ and $\Delta=E_1-E_0$ represents the energy gap of the system. In the third line of Eq.~(\ref{sscheq1}), we have discarded an overall factor. Also, the contributions from higher energy levels have been ignored as they decay much faster than the contribution from the first excited state. From Eq.~(\ref{sscheq1}), it can be seen that the coefficients of the excited states decay exponentially at long times and the characteristic decay time is the correlation time $\zeta_\tau \sim \Delta^{-1}$. As a consequence, for a gapped system, its ground state can be readily found.\cite{vidal1,vidal,jiang,justin} Unfortunately, when a system approaches its critical point, $\Delta\rightarrow 0$ and thus $\zeta_\tau$ tends to infinity. This is the critical slowing down occurring with critical behavior in quantum phase transitions. Here we are interested in whether or not there exists any universal behavior hidden in the short times.

\section{\label{scaling}Scaling theory of universal imaginary-time quantum critical dynamics}
In this section, we propose a phenomenological scaling theory for SITQCD and apply it to the von Neumann entropy for entanglement in a 1D quantum system. We choose the initial state as a direct product state with a small initial value of order parameter $M_0$. Therefore at $\tau=0$, the correlation length and the correlation time are both zero.

In analogy with the classical situation,\cite{Janssen,Zheng} we suggest that after a transient period of a microscopic time scale $\tau_{\rm mic}$, the scaling transformation of the order parameter $M$ near a quantum critical point is given by
\begin{equation}
M(\tau,g,M_0)=b^{-\beta/\nu}M(\tau b^{-z},gb^{1/\nu},M_0b^{x_0}) \label{op}
\end{equation}
for a rescaling of factor $b$, where $g$ is the distance to the quantum critical point and $x_0$ is a new exponent to be related to the initial exponent $\theta$ below. We have neglected dimensional factors for simplicity. Equation (\ref{op}) is assumed to describe the universal behavior of the order parameter in both long times and short times after $\tau_{\rm mic}$.

To see the initial slip, one chooses $b=\tau^{1/z}$ and Eq.~(\ref{op}) becomes
\begin{equation}
M(\tau,g,M_0)=\tau^{-\beta/\nu z}f_M(g\tau^{1/\nu z},M_0\tau^{x_0/z}), \label{op1}
\end{equation}
where $f_M$ is a scaling function related to $M$ (similar definitions will always be implied). For $g=0$ and small $M_0\tau^{x_0/z}$, we expand the right hand side of Eq.~(\ref{op1}) in its second argument and find
\begin{equation}
M(\tau,M_0)=M_0\tau^\theta f^{'}_M(0,0)+\tau^{-\beta/\nu z}O((M_0 \tau^{x_0/z})^3) \label{op2}
\end{equation}
with $\theta$ satisfying a scaling law
\begin{equation}
x_0=\theta z+\beta/\nu \label{op3}
\end{equation}
similar to the classical case,\cite{Janssen} where the prime denotes a partial derivative with an argument.
In (\ref{op2}), we have dropped even order terms because $M$ must have identical sign with $M_0$ and so is an odd function of $M_0$. From Eq.~(\ref{op2}), one sees that when $\tau$ is small, $M(\tau,M_0)\propto M_0\tau^\theta$. This is the critical initial slip in which $M$ increases with $\tau$. Near $\tau_{\rm cr}\sim M_0^{-z/x_0}$, there happens a crossover from the initial slip to the power-law decay stage in which $M\sim \tau^{-\beta/\nu z}$,\cite{Janssen} because at late times the initial condition becomes irrelevant and so does the related argument. Note that $\tau_{\rm cr}$ decreases as $M_0$ increases similar to the classical case.\cite{Janssen,Li3}

The effect of deviations from the critical point can be taken into account. For $g\neq0$, $f_M'$ is now a function of $g\tau^{1/\nu z}$. So, if this argument is small, i.e., $\tau\ll\zeta$, which means that the (imaginary) time is shorter than the correlation time, we can expand in it and get
\begin{equation}
M\simeq M_0\tau^\theta f^{'}_M(0,0)+\Delta M \label{opig}
\end{equation}
with
\begin{equation}
\Delta M\equiv \tau^{\theta+1/\nu z}M_0 gf^{''}_M(0,0) \label{dopig}
\end{equation}
being the leading contribution from the finite $g$, where Eq.~(\ref{op3}) has been used. The reason for this cross term is that if $M_0=0$, $M$ will keeps zero in both the paramagnetic and the ferromagnetic phases after a quench starting from the paramagnetic phase. Equation (\ref{dopig}) shows that for $g\neq 0$, $\Delta M$ deviates from $M_0\tau^\theta f^{'}_M$ to different directions depending on the sign of $g$.

We can of course choose different rescaling factors and obtain different scaling forms from Eq.~(\ref{op}) such as
\begin{equation}
M(\tau,g,M_0)=g^\beta f_{M1}(\tau^{-1} g^{-\nu z},M_0g^{-\nu x_0}), \label{opex}
\end{equation}
and
\begin{equation}
M(\tau,g,M_0)=M_0^{\beta/\nu x_0} f_{M2}(\tau M_0^{z/x_0},gM_0^{-1/\nu x_0}), \label{opm0}
\end{equation}
where $f_{M1}(X,Y)=X^{-\beta/\nu z}f_M(X^{1/\nu z},X^{x_0/z}Y)$ and $f_{M2}(X,Y)=X^{-\beta/\nu z}f_M(X^{1/\nu z}Y,X^{x_0/z})$. The initial slip appears when $gM_0^{-1/\nu x_0}\ll1$ as well as $\tau g^{\nu z}\ll1$ and $\tau M_0^{z/x_0}\ll1$, while the long-time decay occurs in the other limits. Crossovers happen near $g_{\rm cr}\sim M_0^{1/\nu x_0}$, $\tau_{\rm cr}(g)\sim g^{-\nu z}\sim\zeta$ as well as $\tau_{\rm cr}$, respectively. These scaling forms all describe the same scaling behavior and any one can be applied to study both short-time and long-time behaviors as they are related to each other.

A peculiar property in quantum criticality is the entanglement near the critical point. Entanglement is usually measured by the von Neumann entropy, which is defined by $S=-\textrm{Tr}(\rho\textrm{log}\rho)$, where $\rho$ is the reduced density matrix of half of the system.\cite{Eisert,Amico,osterloh} For a 1D system near its critical point, $S=(c/6)\textrm{log}\xi$, where $c$ is the central charge.\cite{Eisert,Amico,osterloh} This property is shared by both symmetry-breaking phase transitions and topological phase transitions.\cite{Eisert,Amico} For the universal short imaginary time evolution, substituting $\xi=\tau^{1/z}f_\xi(g\tau^{1/\nu z},M_0\tau^{x_0/z})$ into $S$ leads to
\begin{equation}
S(\tau,g,M_0)=\frac{c}{6z}\textrm{log}\tau+\frac{c}{6}\textrm{log}f_\xi(g\tau^{1/\nu z},M_0\tau^{x_0/z}). \label{ee}
\end{equation}
Thus we have
\begin{eqnarray}
\begin{split}
\Delta S(\tau,g,M_0)&\equiv S(\tau,g,M_0)-S(\tau,0,0)\\&=f_{S}(g\tau^{1/\nu z},M_0\tau^{x_0/z}), \label{dds}
\end{split}
\end{eqnarray}
where
\begin{equation}
f_{S}=\frac{c}{6}\textrm{log}\frac{f_\xi(g\tau^{1/\nu z},M_0\tau^{x_0/z})}{f_\xi(0,0)}. \label{fxi}
\end{equation}
At the critical point $g=0$, upon expanding the scaling function $f_{S}$ in $M_0\tau^{x_0/z}$, $\Delta S$ becomes
\begin{equation}
\Delta S(\tau,0,M_0)\propto M_0^2\tau^{2x_0/z}\label{ds}
\end{equation}
where odd order terms equal to zero because $\xi$ arises from the correlation function that includes two $M$ and so is an even function of $M_0$.

Similar to the order parameter, we can write the scaling form~(\ref{fxi}) in other forms. For example,
\begin{equation}
\Delta S=f_{S1}(g\tau^{1/\nu z},M_0g^{-x_0\nu}). \label{dsex}
\end{equation}
This enables us to further check the scaling of the entanglement entropy.

\section{\label{theta}Verification of universal imaginary-time quantum critical dynamics}
In this section, we shall confirm the scaling theory proposed for the universal imaginary-time quantum critical dynamics and determine the initial-slip exponent and $x_0$. Both a mean-field theory and numerical results will be presented. We find that although mean-field theories have been reported to be able to explain some experimental results in short real-time dynamics,\cite{Lamacraft} our mean-field theory does not predict the imaginary-time initial slip in disagreement with the numerical results.

\subsection{Model, numerical method and initial state}
We take the $1$D transverse field Ising model as an example. The Hamiltonian is
\begin{equation}
H_I=-\sum\limits_{n=1}^{N-1}\sigma_n^z\sigma_{n+1}^z-h_x\sum\limits_{n=1}^N\sigma_n
^x, \label{HIsing}
\end{equation}
where $\sigma_n^x$ and $\sigma_n^z$ are the Pauli matrices in $x$ and $z$ direction, respectively, at site $n$ and $h_x$ is the transverse field. We have set the Ising
coupling to unity as our energy unit. The critical point of Model~(\ref{HIsing}) is $h_{xc}=1$, the exact critical exponents $\beta=1/8$, $\nu=1$, and $z=1$,\cite{sachdev} and the central charge $c=1/2$.\cite{Eisert,Amico} The order parameter is defined as $M=(1/N)\sum_{n=1}^N\langle\sigma^z_n\rangle$, where $N$ is the total number of spins. This model is realized in CoNb$_2$O$_6$ experimentally.\cite{Coldea}

In order to show the universality of $\theta$, we also use the quantum Ising ladder.\cite{scholl} The Hamiltonian is
\begin{eqnarray}
\begin{split}
H_L=&-\sum\limits_{n=1}^{N-1}\sum\limits_{\alpha=1}^{2}\sigma_{\alpha,n}^z\sigma_{\alpha,n+1}^z-\sum\limits_{n=1}^{N}\sigma_{1,n}^z\sigma_{2,n}^z \\&-h_x\sum\limits_{n=1}^N\sum\limits_{\alpha=1}^{2}\sigma_{\alpha,n}
^x, \label{HIsingl}
\end{split}
\end{eqnarray}
where the first and the second terms are the interactions along the ladder and on the rung, respectively, the third term is the transverse field contribution, and $\alpha$ denotes the two legs of the ladder. The critical point of this model has been determined by the finite-time scaling method~\cite{Yin} to be $h_x=1.8323$ \cite{chenzy} and the critical exponents determined by the same method show that it belongs to the universality class of Model (\ref{HIsing}).\cite{chenzy}

We use the infinite time-evolving block-decimation (iTEBD) algorithm~\cite{vidal} to calculate the imaginary time evolution. As a variant of a time-dependent density-matrix renormalization group, the iTEBD algorithm represents a state in a matrix product form and every site is attached by such a matrix. These matrices are updated according to the local evolution operators, which is obtained by the Suzuki-Trotter decomposition of $\exp(-H \tau)$. When a system is translational invariant, only the matrices in a primitive cell need to be considered. Thus the iTEBD method can simulate an infinite-size lattice efficiently. Errors are induced by accumulation of the errors in the time discretization and the truncations of singular values in every Suzuki-Trotter expansion step. The time interval is chosen as $0.01$.  This time interval is chosen by a compromise between these two kinds of errors. Apparently, a smaller time interval will decrease the errors from the discretization, but increases the errors from the accumulation of the truncations since more steps are needed. For Model~(\ref{HIsing}), we keep $100$ states. We have tested that for more kept states and smaller time intervals, the results have no appreciable changes. Although the fitting error in our calculation is tiny as we shall see, three decimal places are kept in our results from fitting. More accurate results are expected if the algorithm is improved by including the complex canonicalization process to reduce the truncation error. \cite{vidalcan}

The initial state with an order parameter $M_0$ is chosen as a direct product state
\begin{equation}
|\psi_0\rangle=\bigotimes\limits_{n} \left[(a_{2n} |\uparrow\rangle+b_{2n} |\downarrow\rangle)(a_{2n+1} |\uparrow\rangle+b_{2n+1} |\downarrow\rangle)\right] \label{istate},
\end{equation}
where $a_n$ and $b_n$ are the coefficients of the local state at site $n$, $|\uparrow\rangle$ and $|\downarrow\rangle$ are eigenvectors of $\sigma^z$. This state has been factorized into paired terms for convenience of the iTEBD algorithm. \cite{vidal} Two kinds of the state are chosen. One is a homogeneous state, in which $a_{2n}=a_{2n+1}=\sqrt{(1+M_0)/2}$ and $b_{2n}=b_{2n+1}=\sqrt{(1-M_0)/2}$ for a given $M_0$. The other is a staggered state, in which $a_{2n}=\sqrt{(1+M_{0A})/2}$, $b_{2n}=\sqrt{(1-M_{0A})/2}$, $a_{2n+1}=\sqrt{(1+M_{0B})/2}$ and $b_{2n+1}=\sqrt{(1-M_{0B})/2}$ with $M_0=(M_{0A}+M_{0B})/2$, where $M_{0A}$ and $M_{0B}$ are the magnetizations for the even and the odd sublattices, respectively. We shall show that universal critical behavior in the short times is not sensitive to the specific choice of the coefficients. Therefore the homogeneous initial state will be selected except explicitly stated otherwise. In addition, we have all chosen the real variables in the wave function. We shall find in the following that the angular part does not affect the universal behavior.

\subsection{Mean-field theory}
In this section, we shall study a mean-field theory of Model~(\ref{HIsing}). We shall see that this theory satisfies the scaling theory in Sec.~\ref{scaling} with mean-field exponents.
\subsubsection{Mean-field Hamiltonian and its static properties}
The Hamiltonian per site $\tilde{H}_{MF}$ for the quantum Ising model (\ref{HIsing}) in the mean-field approximation is\cite{Continentino,Chakrabarti}
\begin{equation}
\tilde{H}_{MF}=-2M\sigma^z-h_x\sigma^x, \label{mfH}
\end{equation}
where the magnetization $M$ is
\begin{equation}
M=\frac{\langle\psi^{MF}|\sigma^z |\psi^{MF} \rangle}{\langle\psi^{MF}|\psi^{MF} \rangle}, \label{mfopdef}
\end{equation}
in which $MF$ indicates variables in the mean-field approximation. The ground state of the Hamiltonian (\ref{mfH}) is
\begin{equation}
|\psi^{MF} \rangle_G={Z_{MF}}\left(
                                       \begin{array}{c}
                                         \left(2M+\sqrt{h_x^2+4M^2}\right)/h_x \\
                                         1 \\
                                       \end{array}
                                     \right)
, \label{sground}
\end{equation}
where
\begin{equation}
Z_{MF}=\frac{h_x}{\sqrt{8M^2+2h_x^2+4M\sqrt{h_x^2+4M^2}}}.\label{zsground}
\end{equation}
From Eqs. (\ref{mfopdef}), (\ref{sground}) and (\ref{zsground}), one obtains a self-consistent equation for $M$,
\begin{equation}
M=\frac{2M}{\sqrt{h_x^2+4M^2}}.\label{mground}
\end{equation}
One solution, $M=0$, corresponds to the paramagnetic phase, while other solutions,
\begin{equation}
M=\pm \frac{\sqrt{4-h_x^2}}{2}, \label{mground1}
\end{equation}
correspond to the ferromagnetic phase. The critical point at which the two phases coincide is $h^{MF}_{xc}=2$,\cite{Continentino,Chakrabarti} apparently larger than the exact one $h_{xc}=1$. Defining $g_{MF}\equiv h_x-h^{MF}_{xc}=h_x-2$, one obtains $\beta^{MF}=1/2$ from Eq. (\ref{mground1}) and $M\sim (-g_{MF})^\beta$.

\subsubsection{Mean-field dynamics and its scaling behavior}
The evolution of $M$ can be found by expressing $|\psi^{MF} \rangle$ explicitly as $|\psi^{MF} \rangle \equiv(\psi_u-\textrm{i}\varphi_u,\psi_d-\textrm{i}\varphi_d)^{\dagger}$ in the basis of $\sigma^z$. As a result, $M=(\psi_u^2+\varphi_u^2-\psi_d^2-\varphi_d^2)/(\psi_u^2+\varphi_u^2+\psi_d^2+\varphi_d^2)$. Using Eqs.~(\ref{mfH}) and (\ref{scheq}), one then obtains
\begin{equation}
\frac{d M}{d \tau}=4M-4M^{3}-2h_xM\frac{\psi_u\psi_d+\varphi_u\varphi_d}{\psi_u^2+\varphi_u^2+\psi_d^2+\varphi_d^2}. \label{cmfM}
\end{equation}
By parameterizing the real variables of the wave function in the polar coordinates, one finds
\begin{equation}
\frac{\psi_u\psi_d+\varphi_u\varphi_d}{\psi_u^2+\varphi_u^2+\psi_d^2+\varphi_d^2}=\frac{\sqrt{1-M^2}}{2}\cos\Phi, \label{cmfM1}
\end{equation}
where $\Phi$ represents the phase difference between $(\psi_u+\textrm{i}\varphi_u)$ and ($\psi_d+\textrm{i}\varphi_d$). $\cos\Phi$ follows the dynamic equation
\begin{equation}
\frac{d \cos\Phi}{d \tau}=\frac{2h_x(1-\cos^2\Phi)}{\sqrt{1-M^2}} \label{costhe}
\end{equation}
from Eqs.~(\ref{mfH}), (\ref{scheq}), and (\ref{cmfM}). For small $M$, we neglect the square root in Eq.~(\ref{costhe}) and find
\begin{equation}
\tan\left[\Phi(\tau)/2\right]\simeq C\textrm{exp}(-2h_x\tau). \label{scosthe}
\end{equation}
where $C=\tan[\Phi(0)/2]$. So, after a transient time of $1/2h_x$, $\tan\Phi\rightarrow0$ and $\Phi\rightarrow0$ and the phase difference dies out and does not affect the universal behavior.
Therefore, we can choose $\varphi_u=\varphi_d=0$ and so $\cos\Phi=1$ and the evolution equation for $M$ becomes
\begin{equation}
\frac{d M}{d \tau}=4M-4M^{3}-2h_xM\sqrt{1-M^{2}}, \label{mfM}
\end{equation}
which, for small $M$, reads
\begin{equation}
\frac{d M}{d \tau}=-2g_{MF}M-(4-h_x)M^{3}. \label{mfM4}
\end{equation}

Equation~(\ref{mfM4}) is the familiar uniform Ginzburg-Landau theory. It can be readily solved analytically. For $h_x=h^{MF}_{xc}=2$ or $g_{MF}=0$, the solution is
\begin{equation}
M(\tau,M_0)= \textrm{sgn}(M_0)\tau^{-1/2}\sqrt\frac{M_0^2 \tau}{1+4M_0^2 \tau}, \label{mfsM}
\end{equation}
where $\textrm{sgn}(M_0)$ is the sign function, indicating that $M$ has identical sign with $M_0$.
For $\tau\rightarrow\infty$, $M$ approaches zero as $M\simeq \textrm{sgn}(M_0)/(2\tau^{1/2})$. This shows that $M$ depends only on the sign of $M_0$, but not on the magnitude of $M_0$. Also, since in the long-time stage, $M\sim \tau^{-\beta/\nu z}$, substituting $\beta^{MF}=1/2$ we find
\begin{equation}
\nu^{MF}z^{MF}=1.\label{nzmf}
\end{equation}
In fact, Eq.~(\ref{mfsM}) is just in the form of Eq.~(\ref{op1}) with
\begin{equation}
x_0^{MF}/z^{MF}=1/2,\label{x0zmf}
\end{equation}
which leads to $\theta^{MF}=0$ from Eqs.~(\ref{op3}) and (\ref{nzmf}). This can also be obtained by comparing Eq.~(\ref{op2}) with
\begin{equation}
M(\tau,M_0)\simeq M_0-2\tau M_0^3, \label{mfssM}
\end{equation}
which is the short-time and small $M_0$ approximation of Eq.~(\ref{mfsM}). Equations~(\ref{mfsM}) and (\ref{mfssM}) indicate that $M$ will decrease as $\tau$ increases as no initial slip ($\theta^{MF}=0$) appears in the mean-field approximation. However, the whole imaginary-time evolution at the critical point is universal once Eq.~(\ref{mfM4}) is valid.

For $g_{MF}\neq0$, Eq.~(\ref{mfM}) is solved by
\begin{equation}
M(\tau,M_0)=\frac{\textrm{sgn}(M_0)|g_{MF}|^{1/2}\textrm{e}^{-2g_{MF}\tau}}{\sqrt{ |g_{MF}/M_0^2+(2-h_x/2) (1-\textrm{e}^{-4g_{MF}\tau})|}}, \label{mfsM1}
\end{equation}
which can again be cast into the forms of (\ref{opex}) and (\ref{opm0}) with
\begin{equation}
\nu^{MF} x_0^{MF}=1/2.\label{nx0mf}
\end{equation}

Only two out of the three scaling laws~(\ref{nzmf}), (\ref{x0zmf}), and (\ref{nx0mf}) are independent and the three mean-field exponents cannot be solved out individually. If we assume $\nu^{MF}=1/2$ as usual, we arrive at $x_0^{MF}=1$ and $z^{MF}=2$, which is distinct from the exact one $z=1$.\cite{Continentino,Chakrabarti,mfdynez} On the other hand, if we set $\nu^{MF}=1$ as the exact one, we then have $x_0^{MF}=1/2$ and $z^{MF}=1$. To determine which is the correct one needs consideration of the spatial fluctuations, into which we shall not go as the scaling laws are suffice.

\subsection{Quantum critical initial slip and its qualitative explanation}
\begin{figure}
  \centerline{\epsfig{file=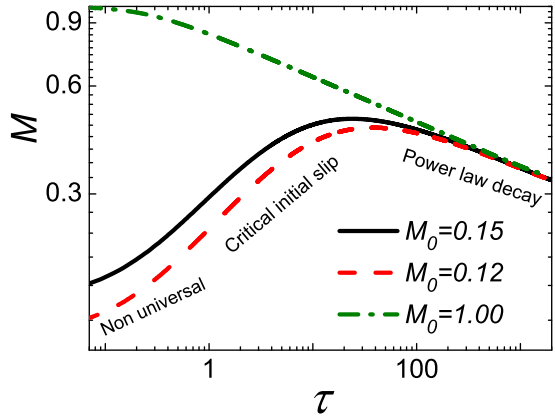,width=1.0\columnwidth}} 
  \caption{\label{region} (color online) Imaginary-time evolution of $M$ for two different small $M_0$ and $M_0=1$ at $h_{xc}$.  Both curves for small $M_0$ show an initial increase for small $\tau$ before a subsequent decay for large $\tau$, confirming the critical initial slip.}
\end{figure}
For the Ising model (\ref{HIsing}), if we start with $M_0=0$, $M$ will keeps zero because of the symmetry restriction. This is different from the classical evolution in which fluctuations can kick it off 0 and all the evolution is initial slip as $\tau_{cr}$ becomes infinity.\cite{Li2} If we start with the saturated order parameter $M_0=1$, $M\sim \tau^{-\beta/\nu z}$ in both short-time and long-time stages, since $M$ cannot increase anymore. In other words, there is no critical initial slip for $M$ starts with $M_0=1$ as shown in Fig.~\ref{region}, which displays the imaginary-time evolution of $M$ at $h_{xc}$. In fact, $M_0=0$ and $M_0=1$ are both fixed points of $M_0$. Indeed, after a small transient, the curve with $M_0=1$ becomes straight with a slope of $0.125$ (the standard deviation of the fit is smaller than $10^{-6}$), which agrees well with $\beta/\nu z=0.125$.

Universal critical initial slip of the order parameter $M$ emerges for small finite $M_0=0$ as seen in Fig.~\ref{region}. After an initial transient stage during which no universal behavior exhibits, one finds the critical initial slip during which lines with nearly identical slope appear showing universality in agreement with Eq.~(\ref{op2}). After passing over its maximum value, $M$ crosses over to the long-time power-law stage in which it decays as $M\sim \tau^{-\beta/\nu z}$ in consistence with the evolution of $M$ from $M_0=1$.  If we regard qualitatively the crossover time $\tau_{\rm cr}$ between the last two stages as the $\tau$ at the peak, $\tau_{\rm cr}$ decreases with the increasing $M_0$, confirming the scaling analysis below Eq.~(\ref{op2}).

\begin{figure}
  \centerline{\epsfig{file=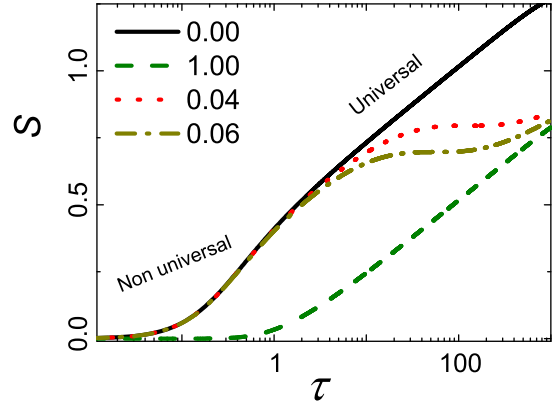,width=1.0\columnwidth}} 
  \caption{\label{regions} (color online) Imaginary-time evolution of $S$ at $h_{xc}$ for different $M_0$ indicated. Note the semi-logarithmic scales.}
\end{figure}
The evolution of the entanglement entropy also shows critical initial slip as can be seen in Fig.~\ref{regions}. $S$ now increases for $M_0=0$ and the evolution is all initial slip after the non-universal transient stage. For $M_0=1$, no initial slip appears and the line reflects long-time behavior. The two curves $S\propto \textrm{log}\tau$ for $M_0=0$ and $M_0=1$ become parallel because both of them are fixed points of $M_0$ and the universal short-time and long-time behaviors are identical. The slopes of the lines are $0.0825$ and $0.0828$, respectively, which are $c/6z$ in agreement with Eq.~(\ref{ee}). The fitting errors are small than $10^{-5}$. The different intercepts result from the different fixed point values of $f_{\xi}$. For small finite $M_0$, $S$ shows the initial slip close to the curve of $M_0=0$ as deviation from the line of $S(\tau,0,0)$ according to Eq.~(\ref{ds}), and then crossover to the long-time behavior along the line of $M_0=1$ since $x_0$ is positive. The crossover happens at a larger $\tau$ for a smaller $M_0$ consistent with the order parameter.

\begin{figure}
  \centerline{\epsfig{file=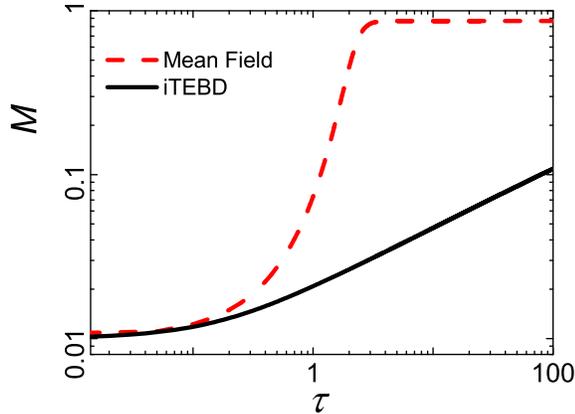,width=1.0\columnwidth}} 
  \caption{\label{mfop} (color online) Imaginary-time evolution of $M$ from the mean-field theory and the iTEBD algorithm at $h_{xc}=1$.}
\end{figure}

The increase of the order parameter $M$ at $g=0$ seems fairly unique in the imaginary time evolution, since according to Eq.~(\ref{sscheq1}), the state of system always approaches to its ground state, whose order parameter is zero at the critical point. Here, the initial vanishing correlation plays an essential role similar to the classical situation.\cite{Li2} In the very early state, there exists only tiny correlation. Consequently, the evolution can be described by the mean-field theory. The critical point determined by the mean field theory, $h^{MF}_{xc}$, is then larger than the exact value $h_{xc}$ due to the suppression of quantum fluctuations.\cite{justin} Therefore, at the real critical point $h_{xc}$, the system ``feels'' a ferromagnetic state of the corresponding mean-field theory in the very early stage and thus the order parameter increases. This also explains the absence of the critical initial slip in the mean-field theory. Indeed, the increase of the order parameter obtained from iTEBD coincides with that obtained from the mean-field theory in the very early transient stage at $h_{xc}=1$ as shown in Fig.~\ref{mfop}. However, deviations appear as the correlation length increases as $\xi\sim \tau^{1/z}+\tau^{1/z}O((M_0\tau^{x_0/z})^2)$ and the effective critical point $h_{xc}^{eff}$ resulted roughly from the fluctuations within $\xi$ decreases from $h^{MF}_{xc}=2$ towards $h_{xc}=1$. Crossover to the long-time decay happens when the accrued $M$ matches that determined by $h_{xc}^{eff}-h_{xc}$.

\subsection{Determination of $\theta$}
\begin{figure}[h]
  \centerline{\epsfig{file=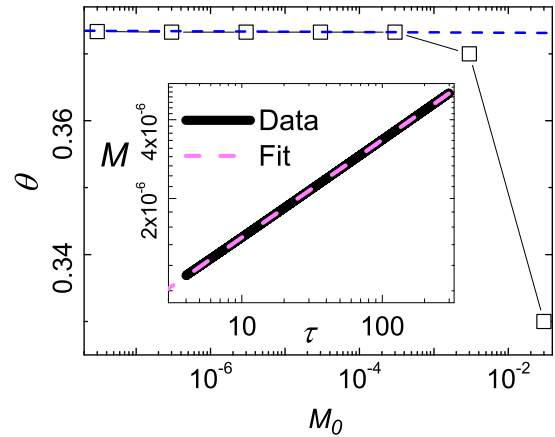,width=1.0\columnwidth}} 
  \caption{\label{iniexp} (color online) $\theta$ estimated for several $M_0$ (squares connected by line). The dashed line is $\theta=0.373$. The inset demonstrates the fitting of $\theta$ for $M_0=3\times 10^{-8}$ at $h_{xc}=1$.}
\end{figure}

\begin{figure}[h]
  \centerline{\epsfig{file=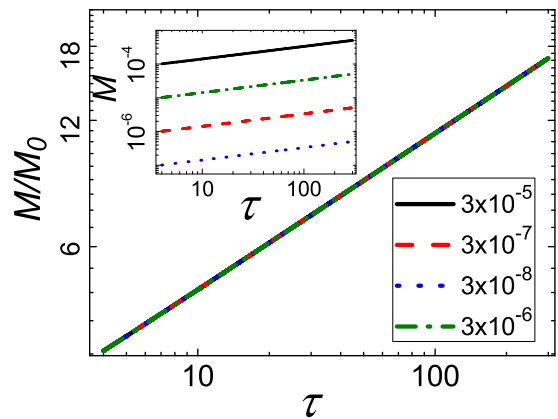,width=1.0\columnwidth}} 
  \caption{\label{iniexp0} (color online) The curves $M$ versus $\tau$ for four $M_0$ in the inset overlap perfectly when $M$ is rescaled with $M_0$.}
\end{figure}

We now determine $\theta$ and the universal properties in the universal short-time stage using the iTEBD method. Figure~\ref{iniexp} shows clearly the universality of $\theta$ obtained for small $M_0$ and thus that of the critical initial slip. The perfect overlap of the linear fit with the numerical results in double logarithmic scales in the inset in Fig.~\ref{iniexp} confirms $M\propto \tau^\theta$. The universal $\theta$ for $M_0<10^{-4}$ is thus $0.373$ with a fitting error of about $10^{-6}$. The fitted $\theta$ becomes smaller when $M_0$ gets larger. This is a result of the higher-order terms in Eq. (\ref{op2}). Figure~\ref{iniexp0} confirms the proportionality of $M$ to $M_0$ at $h_{xc}=1$ because the curves for different $M_0$ collapse perfectly onto each other after rescaling.

\begin{figure}[h]
  \centerline{\epsfig{file=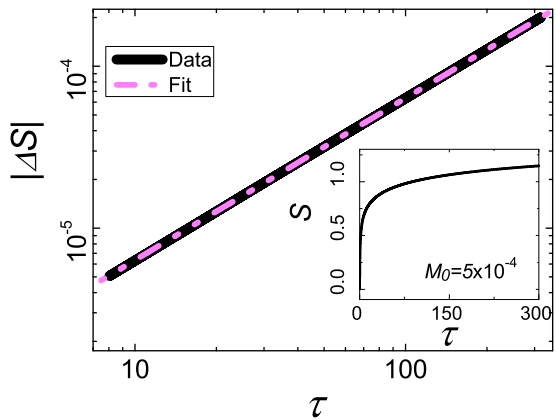,width=1.0\columnwidth}} 
  \caption{\label{iniexp1} (color online) $\Delta S(\tau, 0, 0.0005)$ versus $\tau$. The slope is $2x_0/z=0.998$. The inset shows the imaginary-time evolution $S$ for $M_0=5\times 10^{-4}$ at $h_{xc}=1$.}
\end{figure}

To examine the scaling law (\ref{op3}) and the value of $\theta$ estimated, we show $\Delta S(\tau,0,0.0005)$ in a double logarithmic scale in Fig.~\ref{iniexp1}. According to Eq. (\ref{ds}), the linear fit gives $2x_0/z=0.998$ with a fitting error of $9\times 10^{-6}$. So, $x_0=0.499$ as $z=1$. This value is close to $0.498$ from the scaling law (\ref{op3}) by substituting $\beta$, $\nu$, $z$ and $\theta$.

\begin{figure}[h]
  \centerline{\epsfig{file=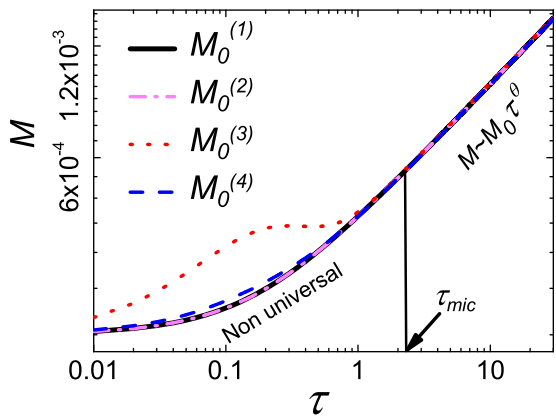,width=1.0\columnwidth}} 
  \caption{\label{depenm} Imaginary-time evolution of $M$ at $h_{xc}=1$ for four different realizations of $M_0$ defined as $M_0^{(1)}$: $a_{2n}=a_{2n+1}=0.50005$ and $b_{2n}=b_{2n+1}=0.49995$ ($M_{0A}=M_{0B}=0.0002$); $M_0^{(2)}$: $a_{2n}=b_{2n}=0.0$, $a_{2n+1}=0.50010$, and $b_{2n+1}=0.49990$ ($M_{0A}=0.0$ and $M_{0B}=0.0004$); $M_0^{(3)}$: $a_{2n}=0.54772$, $b_{2n}=0.44721$, $a_{2n+1}=0.44733$, and $b_{2n+1}=0.54763$ ($M_{0A}=0.2$ and $M_{0B}=-0.1996$) and $M_0^{(4)}$: $a_{2n}=0.50498$, $b_{2n}=0.49497$, $a_{2n+1}=0.49508$, and $b_{2n+1}=0.50488$ ($M_{0A}=0.02$ and $M_{0B}=-0.0196$).}
\end{figure}
Figure \ref{depenm} shows the universality of the scaling behavior with different realizations of the initial state. It can be seen that no matter whether we choose the homogeneous direct product state or the staggered state with the same $M_0$, the scaling functions for $\tau>\tau_{\rm mic}$ are almost identical. This may be understood as follows. In the universal short time region, $\xi\sim \tau^{1/z}+\tau^{1/z}O((M_0\tau^{x_0/z})^2)$. Modes with momentum larger than $1/\xi$ are smeared by the generic quantum critical fluctuation which has an effective momentum $1/\xi$. Thus the initial realization of $M_0$ with effective length scale smaller than $\xi$ will not affect the universal behavior. From the point of view of the renormalization group, the contribution of the modes with large momenta has been integrated out and thus is irrelevant. Note that $M_0$ is not the only ingredient for the critical initial slip. The initial correlation length also plays an important role. If the initial realizations of $M_0$ has a correlation length shorter than $\xi$, they all share identical universal critical initial slip as shown in Fig.~\ref{depenm}. On the other hand, if the initial correlation length is so long that $M_0$ lies in the power decay region, $M$ will continue to decay as $M\sim \tau^{-\beta/\nu z}$ even though the $M_0$ value is identical with the previous one. In this case, no critical initial slip will emerge.

\begin{figure}
  \centerline{\epsfig{file=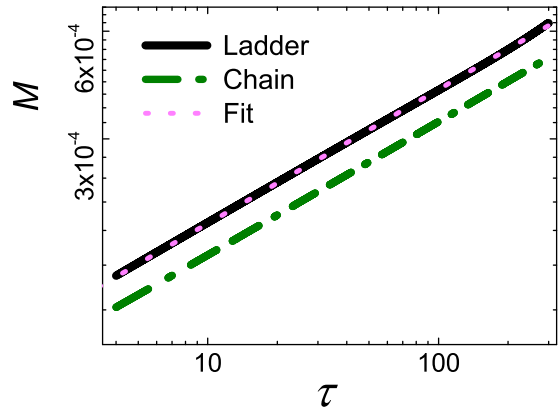,width=1.0\columnwidth}} 
  \caption{\label{unverstheta} (color online) $M$ versus $\tau$ with $M_0=3\times 10^{-5}$ at the critical point of the quantum Ising ladder. $M$ with the same initial condition for the quantum Ising chain is also plotted for comparison. The two parallel curves show an almost identical $\theta$. We have ignored the initial transients.}
\end{figure}

In order to show that the value of $\theta$ only depends on the universality class similar to the classical case,\cite{Janssen,zhengtri} we measure $\theta$ for the quantum Ising ladder, Model (\ref{HIsingl}). Figure \ref{unverstheta} compares the results from both models. The fit gives $\theta=0.374$ with a fitting error of $4\times 10^{-5}$ for Model (\ref{HIsingl}). This value is consistent with that in Model (\ref{HIsing}). The small difference may arise from the accuracy of the critical point determined.

\subsection{Off critical-point effects and crossover to long-time stage}
\begin{figure}
  \centerline{\epsfig{file=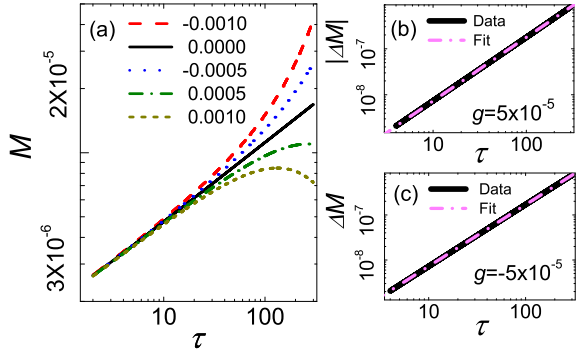,width=1.0\columnwidth}} 
  \caption{\label{finiteg} (color online) (a), $M$ versus $\tau$ for different $g$ with a fixed $M_0=1.0\times10^{-5}$. (b) and (c), fitting $\textrm{log}(\Delta M)$ versus $\textrm{log}\tau$ for $g=\pm5\times10^{-5}$, respectively, to obtain the exponent $\theta+1/\nu z$.}
\end{figure}

In the previous section, we worked at $h_{xc}=1$ to determine the universal short imaginary-time properties. Yet, the scaling forms~(\ref{op1}), (\ref{opig}), and (\ref{opm0}) and those for $S$ can also describe the critical initial slip in the presence of $g$. A direct approximated result from Eq.~(\ref{dopig}) can be seen from Fig.~\ref{finiteg}. The slopes in Fig.~\ref{finiteg}(b) and (c) are fitted as $1.382$ and $1.369$, respectively, with fitting errors smaller than $10^{-4}$. The small differences arise from the higher-order corrections to Eq.~(\ref{dopig}). These results are consistent with $\theta+1/\nu z\simeq 1.372$ by substituting $\theta$, $\nu$, and $z$.

\begin{figure}
  \centerline{\epsfig{file=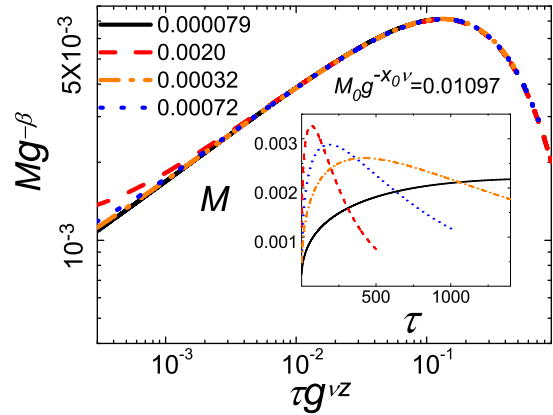,width=1.0\columnwidth}} 
  \caption{\label{fulls} (color online) $Mg^{-\beta}$ versus $\tau g^{\nu z}$ for different $g$ with a fixed $M_0g^{-x_0\nu}$. Inset: $M$ versus $\tau$ for different $g$.}
\end{figure}

Moreover, the scaling forms connect the short-time stage to the long-time stage and describe the universal behaviors in both stages. To see this, Fig.~\ref{fulls} shows the evolution of $M$ for different $g$ with a fixed $M_0g^{-x_0\nu}$ from the non-universal transient stage, passing through the critical initial slip stage, to the decay stage. In the latter two universal stages, the curves in Fig.~\ref{fulls} collapse onto each other after rescaling, confirming Eqs.~(\ref{op2}) and (\ref{opex}).

\begin{figure}
  \centerline{\epsfig{file=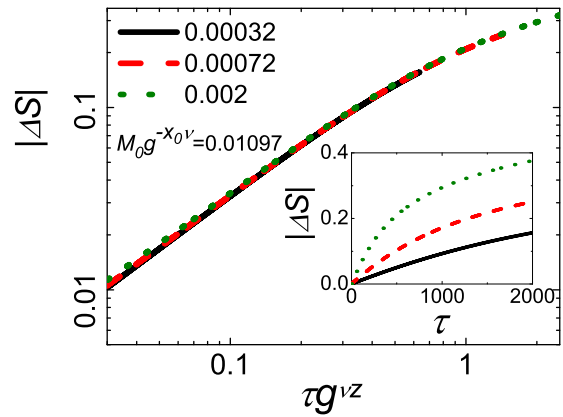,width=1.0\columnwidth}} 
  \caption{\label{fulle} (color online) $\Delta S$ versus $\tau g^{\nu z}$ for different $g$ with a fixed $M_0g^{-x_0\nu}$. Inset: $\Delta S$ versus $\tau$ for different $g$.}
\end{figure}

Similarly, Fig.~\ref{fulle} shows the evolution of $\Delta S$ for different $g$ with a fixed $M_0g^{-x_0\nu}$. In the universal stage in present of finite $g$, curves for different $g$ collapse onto each other after rescaling in agreement with Eqs.~(\ref{dsex}) and (\ref{dds}). In the non-universal stage, the curves deviates from each other slightly.

\subsection{Discussions}
Here we remark on the results. (a) $\theta$ found here is distinctly different from its classical counterpart, which is $0.191(1)$. \cite{Li2,Li3,Zheng1,Zheng,zhengtri,Albano,Grassberger} The reason is that $\theta$ depends on the dynamical equation, which is Eq. (\ref{scheq}) for the quantum dynamics while Langevin's equation for the classical dynamics. (b) Although $\theta$ is different remarkably, $x_0=0.498$ is quite close to its classical counterpart, which is $x_0\simeq 0.539(3)$ using the classical dynamic exponent $z=2.1667(5)$. \cite{Albano,Nightingale} Whether the two $x_0$ should be the same or not is not known at present. However, as both quantum and classical models share identical $\beta/\nu$, it appears likely that both models may share identical $x_0$ too. If this were the case, the quoted classical $z$ would then lead to a classical $\theta=0.172$, about ten percent smaller than the extant value. This is not impossible noticing that in the classical model, the minimum $M_0$ realized in simulation is not very small. \cite{Li3,Zheng1,Zheng,zhengtri,Albano,Grassberger} Note also that increasing $M_0$ reduces $\tau_{\rm cr}$ and thus the time span of initial slip, as can be seen from Fig.~\ref{region}. So, further investigation of the classical model appears desirable. If this $\theta$ were confirmed, it would then imply the conformity of $x_0$ and might be used to estimate the classical $z$. Furthermore, we notice that $x_0$ for the 1D quantum Potts model~\cite{qpotts} is about $0.303$, which is again close to its classical counterpart $x_0=0.285(5)$ using the classical dynamic exponent $z=2.1735(40)$.~\cite{qpottsz} (c) Although $\theta$ is different from the mean field value, $x_0/z=0.498$ is very close to its mean-field value $1/2$. Whether the small difference is genuine or a numerical result needs to be checked further. If they were identical, there would be no universal critical initial slip similar to the present case in the 2D quantum Ising model, because a negative $\theta$ is required from Eq. (\ref{op3}) as $\beta/\nu >0.5$ and $z=1$. \cite{Pelissetto} However, even if they were identical, this would only be a special case. We have checked that for the quantum Potts chain,~\cite{qpotts} similar procedures yield $x_0^{MF}/z^{MF}=1$ from the mean-field theory, while the result from iTEBD gives $x_0/z\simeq 0.303$. Therefore, $\theta$ cannot be generally determined by the known critical exponents. (d) Besides overcoming the critical slowing down, SITQCD has another advantage. In the iTEBD method, the truncation scales as $\textrm{exp}S$. \cite{scholl1} Since $S$ is logarithmically divergent with $\xi$, the necessary truncation should then be infinite in order to obtain accurate critical properties. Otherwise, finite entanglement effects \cite{pollmann} will affect the results. SITQCD thus provides an approach to circumvent this problem to obtain the critical properties at the early stage of evolution at which $S$ is still modest. Indeed, from Fig. \ref{iniexp1}, it can be seen that $S$ increases logarithmically from zero due to the nonentangled initial direct product states. This means we can still take finite truncations in SITQCD. In the following, we shall show how to determine the quantum critical properties with the SITQCD method.

\section{\label{application}Application of short imaginary-time quantum critical dynamics}
We now develop a method based on SITQCD to detect quantum critical properties. We shall first determine the critical properties of the transverse field Ising model. Then we shall show that this method can also be applied to topological quantum phase transitions.

\subsection{Estimating critical properties of quantum Ising model via SITQCD}
According to Eq. (\ref{op1}), it is convenient to fix the term with $M_0$ since it contains an additional initial exponent. As mentioned, $M_0=0$ and $M_0=1$ are both fixed points. For $M_0=0$, the order parameter $M$ maintains zero in the subsequent evolution because $M$ is an odd function of $M_0$ at $g=0$. Accordingly, it is convenient to choose an initial state with $M_0=1$. This resembles the nonequilibrium relaxation critical dynamics. \cite{ito}

To see how to estimate the critical properties, we begin with Eq. (\ref{op1}). For $M_0=1$ and $\tau>\tau_{\rm mic}$, Eq.~(\ref{op1}) is simplified to
\begin{equation}
M(\tau,g)=\tau^{-\beta/\nu z}f_M(g\tau^{1/\nu z}). \label{op4}
\end{equation}
After expanding $f_M(g\tau^{1/\nu z})$ in $g\tau^{1/\nu z}$ for small $g\tau^{1/\nu z}$, we arrive at
\begin{equation}
\textrm{log}M(\tau,g)= -\frac{\beta}{\nu z} \textrm{log}\tau+\textrm{log} f_M(0)+\Delta \textrm{log}M(\tau,g), \label{op5}
\end{equation}
with
\begin{equation}
\Delta \textrm{log}M(\tau,g)=\tau^{1/\nu z}g\frac{f_{M}^{'}(0)}{f_{M}(0)}. \label{op6}
\end{equation}
According to Eqs (\ref{op5}) and (\ref{op6}), at the critical point $g=0$, $\log M(\tau,0)= -[\beta/(\nu z)]\log\tau+\log f_M(0)$; while for $g\neq0$, $\log M(\tau,g)$ deviates from $\log M(\tau,0)$ towards different directions. These then provide a method to fix the critical point. In addition, Eqs.~(\ref{op5}) and (\ref{op6}) can give rise to the exponents.

\begin{figure}
  \centerline{\epsfig{file=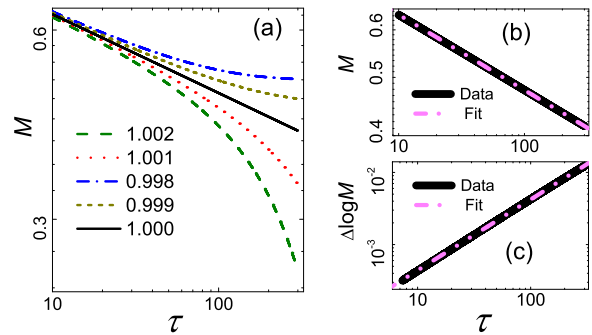,width=1.0\columnwidth}} 
  \caption{\label{detIsCP} (color online) (a) Imaginary-time evolution of $M$ for different $h_x$ in double logarithmic scales. (b) Fitting $\textrm{log}M$ versus $\textrm{log}\tau$ at $h_x=1.000$ to find $\beta/\nu z$. (c) Fitting $\textrm{log}(\Delta \textrm{log}M)$ versus $\textrm{log}\tau$ at $h_x=0.9998$ ($g=0.0002$) to find $1/\nu z$.}
\end{figure}

Figure~\ref{detIsCP} shows that at $h_x=1.000$, the curve of $\textrm{log}M$ versus $\textrm{log}\tau$ is almost straight in the double logarithmic scale; while for $h_x\neq 1.000$, the curves deviate from the straight line. Thus, $h_{xc}=1.000$, consistent with the exact result $h_{xc}=1$. Further, the linear fits in Fig. \ref{detIsCP}(b) and (c) give $\beta/\nu z=0.125$ and $1/\nu z=0.983$ with fitting errors of $7\times 10^{-7}$ and $4\times 10^{-5}$ according to Eqs.~(\ref{op5}) and (\ref{op6}), respectively. Inserting $z=1$, which is obtained by quantum--classical mapping, we get $\beta=0.123$ and $\nu=0.983$. Both are close to their exact values.

\begin{figure}
  \centerline{\epsfig{file=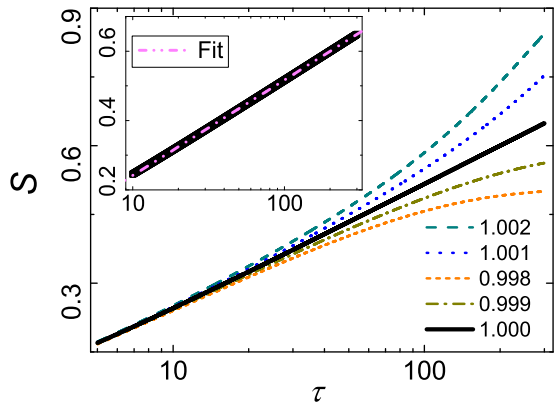,width=1.0\columnwidth}} 
  \caption{\label{detIsCPee} (color online) Imaginary-time evolution of $S$ for different $h_x$ in a semi-logarithmic scale. Inset: fitting $S$ versus $\textrm{log}\tau$ at $h_x=1$ gives $c/z=0.497$.}
\end{figure}

Figure \ref{detIsCPee} shows similar results for $S$ in the universal region. The straight line of $S$ versus $\textrm{log}\tau$ at $h_x=1.000$ gives $c/z=0.497$ with a fitting error of $9\times10^{-6}$ according to Eq.~(\ref{ee}). As $z=1$, we get $c=0.497$, very close to the exact value $c=1/2$. Furthermore, since $S$ is an even function of $M_0$, it can also be calculated with $M_0=0$. Similar results are obtained.

\subsection{Application in topological quantum phase transition}
In the previous section, we determine the critical point and critical exponents of the transverse field Ising model, which exhibits a typical quantum phase transition belonging to the Landau-Ginzburg-Wilson paradigm. The vanishing of the local order parameter, like $M$ in the Ising model, is the signal of the phase transition. But in topological quantum phase transitions, local order parameters cannot be found in principle. In this case, the entanglement entropy $S$ becomes an important quantity to characterize these phase transitions. As we have shown above, $S$ also exhibits universal behavior and contains useful information about phase transitions in SITQCD. Thus we expect that it can also be applied to topological quantum phase transitions. In this section, we apply the SITQCD method to topological phase transition in the anisotropic spin-$1$ Heisenberg model with a single-ion anisotropy. We shall first introduce briefly the model and its equilibrium properties. Then we shall determine its critical properties by the SITQCD method and compare them with the results obtained by other methods.

\subsubsection{Model and its equilibrium critical properties}
The Hamiltonian of the single-ion anisotropic spin-$1$ Heisenberg model in 1D is \cite{scholl,Chen,Albuquerque,Boschi,Ueda,Hu,Tzeng1,Tzeng2,Huang}
\begin{equation}
H_H=\sum\limits_{n=1}^{N-1}\textbf{S}_n\cdot {\textbf{S}}_{n+1}+D\sum\limits_{n=1}^NS^{z2}, \label{Hamil2}
\end{equation}
where $\textbf{S}_n$ is spin-$1$ operator at site $n$ and $D$ stands for the uniaxial single-ion anisotropy. The ground states of~(\ref{Hamil2}) have three phases depending on $D$. For negative $D$, the ground state is the N\'{e}el phase. The Haldane phase appears for relatively larger $D$. Increasing $D$ further, the ground state becomes the large-$D$ phase. Experimentally, this model is realized in some Ni compounds with significant single-ion anisotropies. \cite{ni} According to the symmetry consideration, the N\'{e}el--Haldane transition is the Ising-type transition \cite{Chen} described by a conformal field theory with the central charge $c=1/2$. \cite{Blote} The Haldane--large-$D$ transition is a third order Gaussian transition \cite{Tzeng2} described by a conformal field theory with $c=1$. \cite{Blote} This phase transition is hard to deal with and has attracted a lot of effort. \cite{Chen,Albuquerque,Boschi,Ueda,Hu,Tzeng1,Tzeng2,Huang} Recently an improved density-matrix renormalization-group method has yielded promising results. \cite{Hu} Here we also focus on the Haldane--large-$D$ transition and compare our results mainly with those obtained by this improved density renormalization-group method. \cite{Hu}

The Haldane phase is a typical topological phase in one dimension. There is no local order parameter to characterize this phase. Instead, it is usually described by a nonlocal string order parameter, which can be observed in experiments. \cite{endres} For the model (\ref{Hamil2}), the string order parameter is defined as
\begin{equation}
M_{str}=-\lim_{|j-k|\rightarrow \infty}\left\langle S_j^x \textrm{e}^{i\pi\sum_{n=j+1}^{k-1}S_n^x}S_k^x\right\rangle, \label{strop}
\end{equation}
which is zero in the large-$D$ phase and nonzero in the Haldane phase \cite{scholl,Chen,Albuquerque,Boschi,Ueda,Hu,Tzeng1,Tzeng2,Huang} and can thus characterize the Haldane--large-$D$ transition. In equilibrium, $M_{str}\propto g^{\beta_s}$ near the critical point. \cite{scholl,Chen,Albuquerque,Boschi,Ueda,Hu,Tzeng1,Tzeng2,Huang} Scaling analyses similar to Eq. (\ref{op1}) then implies that at the critical point,
\begin{equation}
M_{str}\propto \tau^{-\beta_s/\nu z},\label{mstr}
\end{equation}
which is similar to the local order parameter in phase transitions belonging to the Landau-Ginzburg-Wilson paradigm. We shall use the scaling behavior of the entanglement entropy to determine the critical point, as it is more universal and easier for calculation. \cite{Tzeng1,Tzeng2} The iTEBD algorithm is also used with a second order Suzuki-Trotter decomposition. The time interval is again $0.01$. $300$ states are kept and the string length $|j-k+1|$ is chosen up to $5000$.

\subsubsection{Estimating critical properties of anisotropic spin-1 Heisenberg Model via SITQCD}
\begin{figure}
  \centerline{\epsfig{file=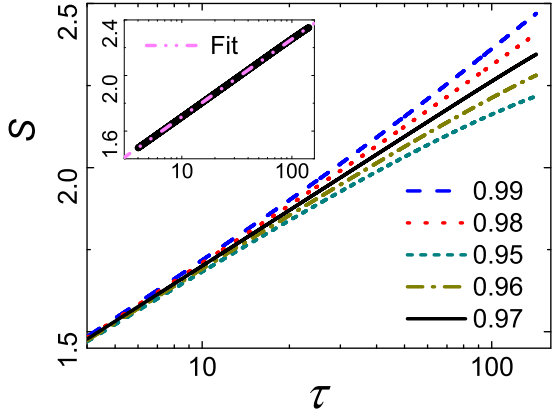,width=1.0\columnwidth}} 
  \caption{\label{eeHL} (color online) Imaginary-time evolution of the $S$ versus $\tau$ for different $D$ in a semi-logarithmic scale. The inset shows the fit at $D=0.97$.}
\end{figure}

Figure \ref{eeHL} shows the evolution of $S$ for several $D$. The straight line gives $D_c=0.97$, which agrees with $D_c=0.96845(8)$ from the improved density-matrix renormalization-group method. \cite{Hu} The precision here is limited by the time span: smaller divisions of $D$ cannot be distinguished within the time span shown in Fig. \ref{eeHL}. The slope of $S$ versus $\textrm{log}\tau$ at $D_c$ gives $c/z=1.001$ with a fitting error $3\times 10^{-5}$ according to Eq.~(\ref{ee}). Thus $c=1.001$, which is close to the exact value $c=1$ since $z=1$.

\begin{figure}
  \centerline{\epsfig{file=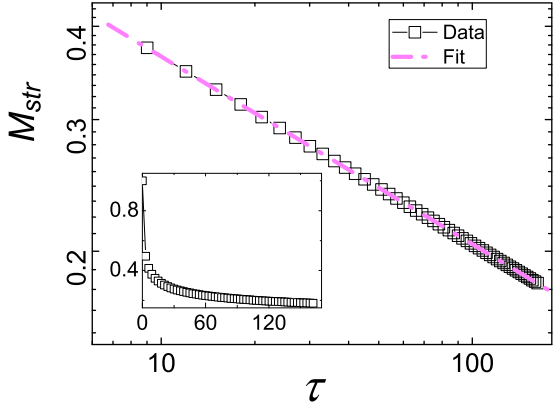,width=1.0\columnwidth}} 
  \caption{\label{opHL} (color online) Imaginary-time evolution of $M_{str}$ in a double logarithmic scale at $D_c$. The line has a slope $\beta_s/\nu z=0.251$. The inset shows the result in linear scales.}
\end{figure}

Figure \ref{opHL} shows the imaginary-time evolution of $M_{str}$. The straight line gives $\beta_s/\nu =0.251$ with a fitting error $8\times 10^{-5}$. This result is consistent with $\beta_s/\nu=0.239$ predicted by the density-matrix renormalization-group method. \cite{Hu}

\section{\label{real}Short real-time dynamics}
We have studied the universal short-time dynamics with a direct product state in imaginary time. However, experimental implementation and observation are both in real time. Therefore, it is worth considering the short real-time dynamics. As we mentioned in Sec. \ref{introduction}, some properties are shared in both real time and imaginary time evolutions. An example is the Kibble-Zurek mechanism. \cite{degrandi2,cwliu,degrandi3} Whether the short-time quantum dynamics can be extended to the real-time situation is explored in this section.

To be explicit, we again consider the 1D quantum Ising model (\ref{HIsing}). The iTEBD algorithm is also used. The time interval is again $0.01$, which is identical to that in the imaginary situation; while the number of states is kept to $200$, which has been shown to give reliable results for $t\sim 10$. \cite{vidal} This time span in which the algorithm works well is much smaller than that in the imaginary-time situation. A reason will come out below.

\begin{figure}
  \centerline{\epsfig{file=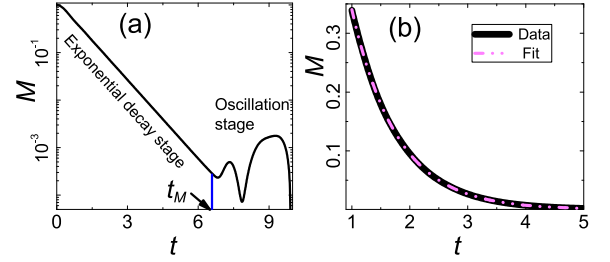,width=1.0\columnwidth}} 
  \caption{\label{rop} (color online) (a) Real-time evolution of $M$ at $h_x=1$. (b) Exponential fit of the evolution at short times.}
\end{figure}

For simplicity, we only consider the initial state with $M_0=1$ and check whether the scaling form~(\ref{op4}) is valid at $g=0$ ($h_x=1$). Figure \ref{rop} shows the time evolution of $M$. Two stages are separated near $t_M\approx 6.5$. When $t<t_M$, $M$ decays exponentially as $M\propto \exp(-t/t_d)$ with a characteristic decay time $t_d$. An exponential fit in Fig.~\ref{rop}(b) yields $t_d\simeq 0.783$ with a fitting error of $3\times10^{-4}$. When $t>t_M$, $M$ oscillates. The behaviors in both $t>t_M$ and $t<t_M$ are apparently different from the imaginary-time evolution. Accordingly, Eq.~(\ref{op4}) cannot describe the real-time dynamics.

A qualitative explanation is as follows. As we know, universal power-law decay of the order parameter is controlled by the low energy levels near the ground state. \cite{sachdev} In the imaginary-time evolution, the system decays quickly to the vicinity of the ground state. Then, its evolution is governed by the low energy levels and exhibits universal power laws. The situation is different for the real-time evolution. Because of the unitary evolution of the real-time dynamics, the excited state will not decay. This may be the reason why $M$ decays exponentially, much faster than the power-law one. For the Kibble-Zurek mechanism applicable in both real-time and imaginary-time situations on the other hand, the initial states are chosen to lie in the vicinity of the ground state. Accordingly, the participation of the excited states should be responsible for losing the universal power-law decay.

\begin{figure}
  \centerline{\epsfig{file=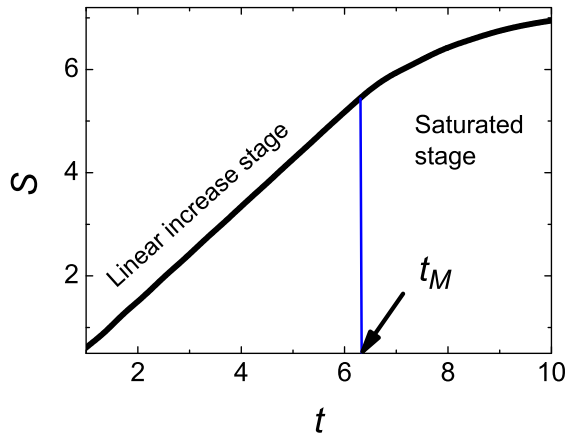,width=1.0\columnwidth}} 
  \caption{\label{rent} (color online) Real-time evolution of $S$ at $h_x=1$.}
\end{figure}

To further support our argument, we also study the evolution of the entanglement entropy $S$. Figure \ref{rent} shows that $S$ increases linearly with $t$ before entering the saturated stage near $t_M$. Similar behavior has been reported. \cite{cardy,Schachenmayer} This is different from the imaginary-time situation, in which $S\propto \textrm{log}\tau$. Since space and time are isotropic in the quantum Ising model, we may assume $\xi\sim t$. This then implies $S\sim \xi$. This indicates that the entanglement entropy $S$ is an extensive quantity similar to the thermal entropy. So, the excited states should dominate the evolution in the short-time stage. In addition, because $S\sim \xi$, the truncation in the iTEBD algorithm should increase exponentially. This may be a reason for invalidating the algorithm in the real-time evolution.

\section{\label{summary}Summary}
This paper focuses on short imaginary-time quantum-critical dynamics with a direct product state as an initial state. Similar to the classical critical phenomena, we have found that there exists a universal critical initial slip at the short imaginary-time stage. This behavior is characterized by a universal exponent $\theta$ for small initial magnetization $M_0$. For the universality class of the 1D quantum Ising model, $\theta=0.373$, which is almost twice of its extant classical counterpart, although the exponent related to $M_0$, $x_0$, is close. In addition, $x_0/z=0.498$ is quite close to its mean-field value $1/2$ for this model. A scaling theory for the universal imaginary-time quantum-critical dynamics during both short times and long times has been proposed and verified both by a mean-field theory and by numerical results. According to the full scaling forms of the short imaginary-time quantum-critical dynamics, the critical point and critical exponents can be effectively determined for either usual quantum phase transitions or topological phase transitions. The short-time method avoids both critical slowing down and a large entanglement entropy that may require large truncations in numerical computations.

\section*{Acknowledgements}

We wish to thank Tao Xiang, Zhiyuan Xie and Anders W. Sandvik for their helpful discussions and Peter Young for his useful comments. We are grateful to an anonymous referee for his/her suggestions of a mean-field approach and the analysis on long-time behavior. This project was supported by NNSFC (10625420) and FRFCUC.


\begin{thebibliography}{99}
\bibitem{mfdynez}Apparently, $z^{MF}$ defined in this nonequilibrium way is different from the quantum-classical mapping which leads to $z=1$. To understand this ambiguous problem, we notice that after the quantum-classical mapping, the quantum Ising model corresponds to a $\phi^4$ field theory with isotropic fluctuations, represented by the gradient terms, in the imaginary-time and space directions. These isotropic gradient terms result in $z=1$ by dimensional analysis. However, when in the mean-field approximation, these fluctuations are neglected totally. Therefore, $z$ has no definition in the mean-field level of the quantum-classical mapping.
\end{thebibliography}
\end{document}